\documentclass[prl,twocolumn,showpacs,aps]{revtex4-1}
\usepackage{color}
\usepackage{amsmath,amssymb}
\usepackage{txfonts}
\usepackage{graphicx}
\usepackage[english]{babel}

\usepackage[unicode]{hyperref}
\hypersetup{
    colorlinks=true,
    linkcolor=blue,
    citecolor=blue,
    filecolor=black,
    urlcolor=blue
}
\def\urlprefix{}
   \def\url#1{}
\usepackage{bm}
\usepackage{epstopdf}
\usepackage{enumerate}

\newcommand{\ve}{\varepsilon}

\newcommand{\ec}{\epsilon_{\rm cut}}
\newcommand{\rC}{\textit{C}}
\newcommand{\rI}{\textit{I}}

\newcommand{\ket}[1]{|#1\rangle}

\newcommand{\lab}[1]{\label{#1}}
\newcommand{\EQ}[1]{\begin{align}#1\end{align}}

\newcommand{\eref}[1]{(\ref{#1})}
\newcommand{\fref}[1]{Fig.~\ref{#1}}

\begin{document}
\title{Brownian motion of a matter-wave bright soliton: realizing a quantum pollen grain}
\author{R. G. McDonald and A. S. Bradley} 
\affiliation{Department of Physics, Centre for Quantum Science, and Dodd-Walls Centre for Photonic and Quantum Technologies, University of Otago, Dunedin 9010, New Zealand.}
\begin{abstract}
Taking an open quantum systems approach, we derive a collective equation of motion for the dynamics of a matter-wave bright soliton moving through a thermal cloud of a distinct atomic species. The reservoir interaction involves energy transfer without particle transfer between the soliton and thermal cloud, thus damping the soliton motion without altering its stability against collapse. We derive a Langevin equation for the soliton centre of mass velocity in the form of an Ornstein-Uhlenbeck process with analytical drift and diffusion coefficients. This collective motion is confirmed by simulations of the full stochastic projected Gross-Pitaevskii equation for the matter-wave field. The system offers a pathway for experimentally observing the elusive energy-damping reservoir interaction, and a clear realization of collective Brownian motion for a mesoscopic superfluid droplet.
\end{abstract}
\maketitle
\begin{figure}[t!]{
\begin{center} 
\includegraphics[width=0.9\columnwidth]{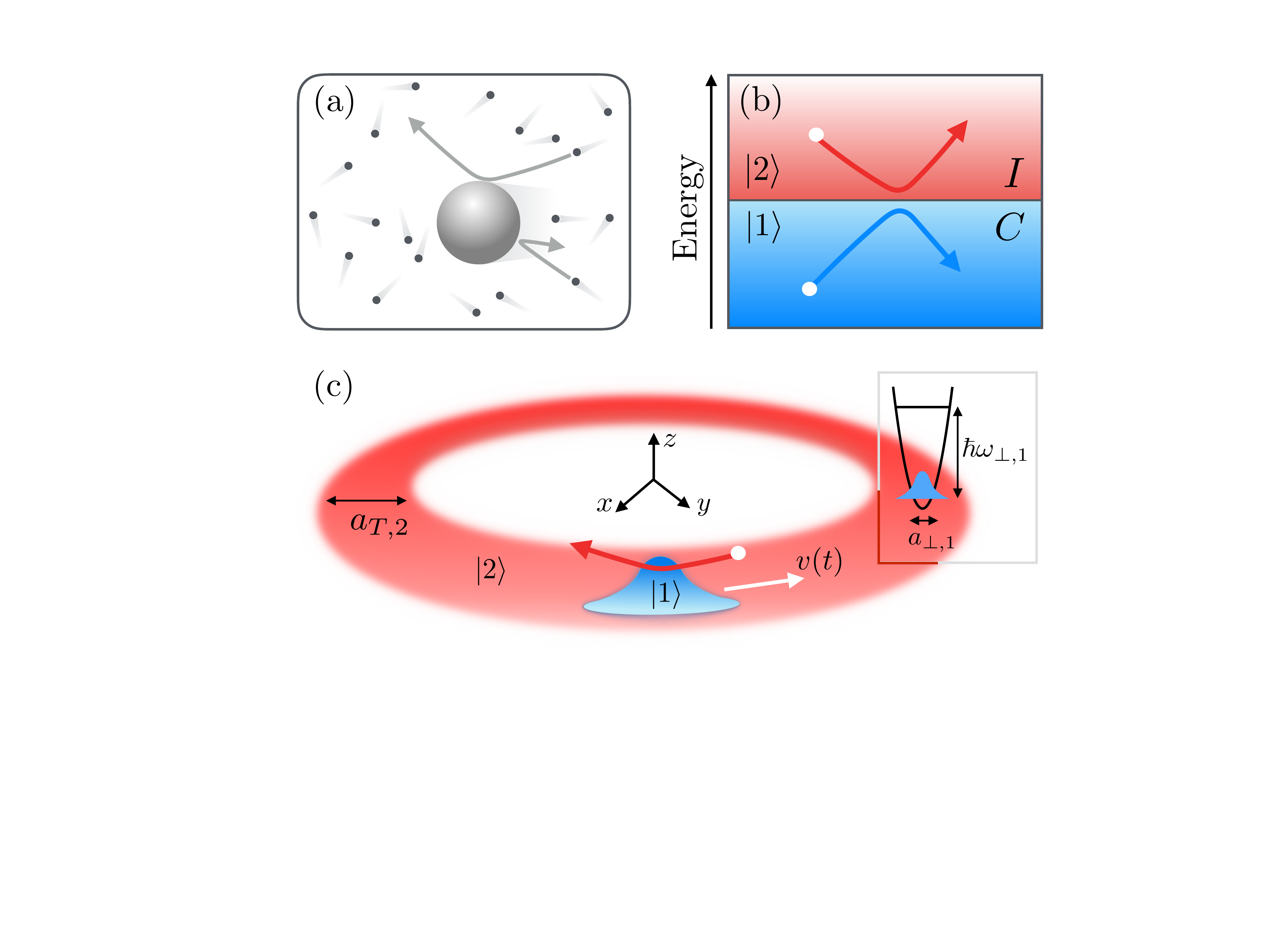}
\caption{(colour online) Schematic: (a) In classical Brownian motion a heavy mass is randomly struck by lighter masses, causing stochastic motion. A bright soliton jostled by thermal atoms of a distinct atomic species provides a matter-wave analogue of Brownian motion --- a ``quantum pollen grain". (b) Representation of the energy-damping reservoir interaction between $\ket{1}$ atoms (coherent, $\rC$ region) and $\ket{2}$ atoms (incoherent, $\rI$ region), in stochastic projected Gross-Pitaevskii theory~\cite{Gardiner:2003bk,Bradley:2014a,Bradley:2015cx}. The interactions are number-conserving, and the thermal cloud of $\ket{2}$ acts as an energy reservoir for the bright soliton of $\ket{1}$. (c) A bright soliton with centre of mass velocity $\varv(t)$, immersed in a distinct thermal cloud. Both species are confined to ring geometries, but with distinct transverse length scales: $a_{\perp,1}\ll a_{T,2}$ [see text].
\label{Fig:schem}}
\end{center}}
\end{figure}

Robert Brown's 1827 observations of jostled pollen grains suspended in water [Fig.~\ref{Fig:schem}~(a)], followed by Einstein's theory of Brownian motion, initiated the study of stochastic dynamics, with deep implications extending from stellar motion to chemical reactions and the quantum jitter of subatomic particles~\cite{Chandrasekhar:1943eb}.
As highly tuneable and ultra-cold quantum systems, atomic Bose-Einstein condensates (BECs) provide a pristine setting for studying Brownian motion in the quantum realm. 
\par
The evolution of open quantum systems depends crucially upon the nature of system-environment interactions. Theories of BEC-reservoir interactions have focused largely on population transfer~\cite{Choi:1998eh,Zaremba:1999iu,Stoof:1999tz,Tsubota2002} as the process driving dissipation. This \emph{number-damping} process plays a central role in condensate growth during evaporative cooling~\cite{QKIII,QKPRLII,Anglin:1999fn,Weiler:2008eu}, dissipation of excitations such as collective modes~\cite{Morgan:2003bv}, solitons~\cite{Cockburn10a} and vortices~\cite{Rooney:2010dp}, the formation of vortex lattices~\cite{Penckwitt2002}, and spontaneous defect formation during a quench~\cite{Weiler:2008eu,Su:2013dh}. However, an additional process of fundamental importance causes dissipation \emph{without} population transfer~\cite{Gardiner:2003bk,Rooney:2012gb,McDonald:2015ju}. This  \emph{energy-damping} reservoir interaction~\cite{QKVI} is essential in sympathetic cooling~\cite{Myatt:1997ct}, drives superfluid internal convection~\cite{Gilz:2011jma}, has an analogue in inelastic light scattering~\cite{Daley:2014hc}, and may also underpin anomalous energy damping in a spinor BEC~\cite{Liu:2009en}, and play a dominant role in vortex decay~\cite{Rooney:2016wr}. Yet a clear experimental observation of energy-damping has remained elusive. 
\par
As localised waves that propagate with a permanent functional form, solitons \cite{Drazin:1989ub} appear as solutions of a large class of partial differential equations that are both dispersive and weakly nonlinear. They have been observed in a range of systems including water waves \cite{Russell:1845tw,Chabchoub:2013kc}, temporal~\cite{Emplit:1987ht} and spatial~\cite{Bjorkholm:1974dz} optical pulses, BECs \cite{Burger:1999ew,Strecker2002a,Khaykovich:2002dk,Hamner11a}, and Fermi superfluids~\cite{Yefsah:2013bl}. They have also been used to characterize critical dynamics~\cite{Zurek:2009hl,Lamporesi:2013bi,Damski10a}, and as robust wavepackets for matter-wave interferometry~\cite{McDonald:2014fg}. Provided the geometry is sufficiently prolate~\cite{Carr:2000bq}, the bright soliton is an analytic solution of the 1D Gross-Pitaevskii equation describing dynamics of zero temperature BECs with attractive interactions~\cite{Carr:2000bw}. 
As particle-like nonlinear excitations sensitive to thermal and quantum fluctuations~\cite{Dziarmaga04a,Sinha:2006hi,Martin2010a,Martin2010b,Cockburn10a}, bright solitons provide a unique probe of reservoir interactions.
\par
In this Letter we study the motion of a bright soliton through a thermal cloud of a distinct atomic species. Recent theoretical work has extended the complete stochastic projected Gross-Pitaevskii equation (SPGPE)~\cite{Gardiner:2003bk,Rooney:2012gb} to multicomponent systems~\cite{Bradley:2014a}, and to an effective 1D description of prolate systems~\cite{Bradley:2015cx}, allowing a rigorous formulation of the bright soliton motion in terms of an SPGPE for the dynamics of the soliton matter-wave field. 
The system we consider consists of a two-component mixture of Bose gases, $\ket{1}$ and $\ket{2}$, with similar~\footnote{The theory developed in \cite{Bradley:2014a} gives an exact semi-classical treatment of the reservoir coupling rates for equal masses, providing a good approximation for nearly equal masses; the rates for arbitrary masses will be addressed in future work.} constituent masses~[Fig.~\ref{Fig:schem} (b)]. We consider the regime where $\ket{1}$ is Bose-condensed with negligible thermal fraction, while $\ket{2}$ is noncondensed, i.e. $T_{c,2}\lesssim T\ll T_{c,1}$, where $T_{c,i}$ is the critical temperature for Bose-Einstein condensation of $\ket{i}$. In language of c-field theory~\cite{Blakie:2008isa}, $\ket{1}$ and $\ket{2}$ form coherent ($C$) and incoherent ($I$) fields respectively. We consider a geometry regime where $\ket{1}$ is restricted to one-dimensional motion, while $\ket{2}$ maintains three dimensional characteristics, as shown schematically in Fig.~\ref{Fig:schem} (c). This is a regime of buffer gas cooling, realizable for two hyperfine states of the same atom via magic wavelength techniques \cite{Safronova:2012jg}, or for distinct atoms by optical control of the transverse potentials. In a strict quasi-1D regime for both components, the reservoir interactions are greatly complicated by the modified dispersion. The advantage of considering an embedded 1D regime is that it allows for effective 1D superfluid dynamics, whilst preserving the simpler form of 3D reservoir interactions~\cite{Bradley:2015cx}.
\par
We take as our starting point the two-component 3D SPGPE derived in~\cite{Bradley:2014a}, and impose confinement geometry giving effective 1D superfluid motion immersed in a 3D thermal cloud, allowing an effective 1D description of our open quantum system~\cite{Bradley:2015cx}. In what follows we neglect the dynamics of $\ket{2}$, a reasonable approximation for the regime of tight transverse confinement~\footnote{The 3D thermal cloud of {$\ket{2}$} will equilibrate quickly relative to the timescale for 1D evolution of the superfluid in {$\ket{1}$}.}. Relative to the one-component system, the distinguishable reservoir introduces two changes to the theory. Firstly, provided the BEC is sufficiently cold, i.e. $T\ll T_{c,1}$, the reservoir interactions between the $C$-region and $I$-region of $\ket{1}$ are unimportant. Secondly, the only reservoir interaction between the $C$-region of $\ket{1}$ and the $I$-region of $\ket{2}$ takes a form similar to one-component energy damping~\cite{Rooney:2012gb}, modified due to the distinguishable nature of the $s$-wave interaction with scattering length $a_{12}$~\cite{Bradley:2014a}. In this regime an exact SPGPE for the dynamics of $\ket{1}$ is found via the theory developed in~\cite{Gardiner:2003bk,Bradley:2014a,Bradley:2015cx} as the Langevin equation in Stratonovich form
\begin{figure}[tb!]{
\begin{center} 
\includegraphics[width=0.8\columnwidth]{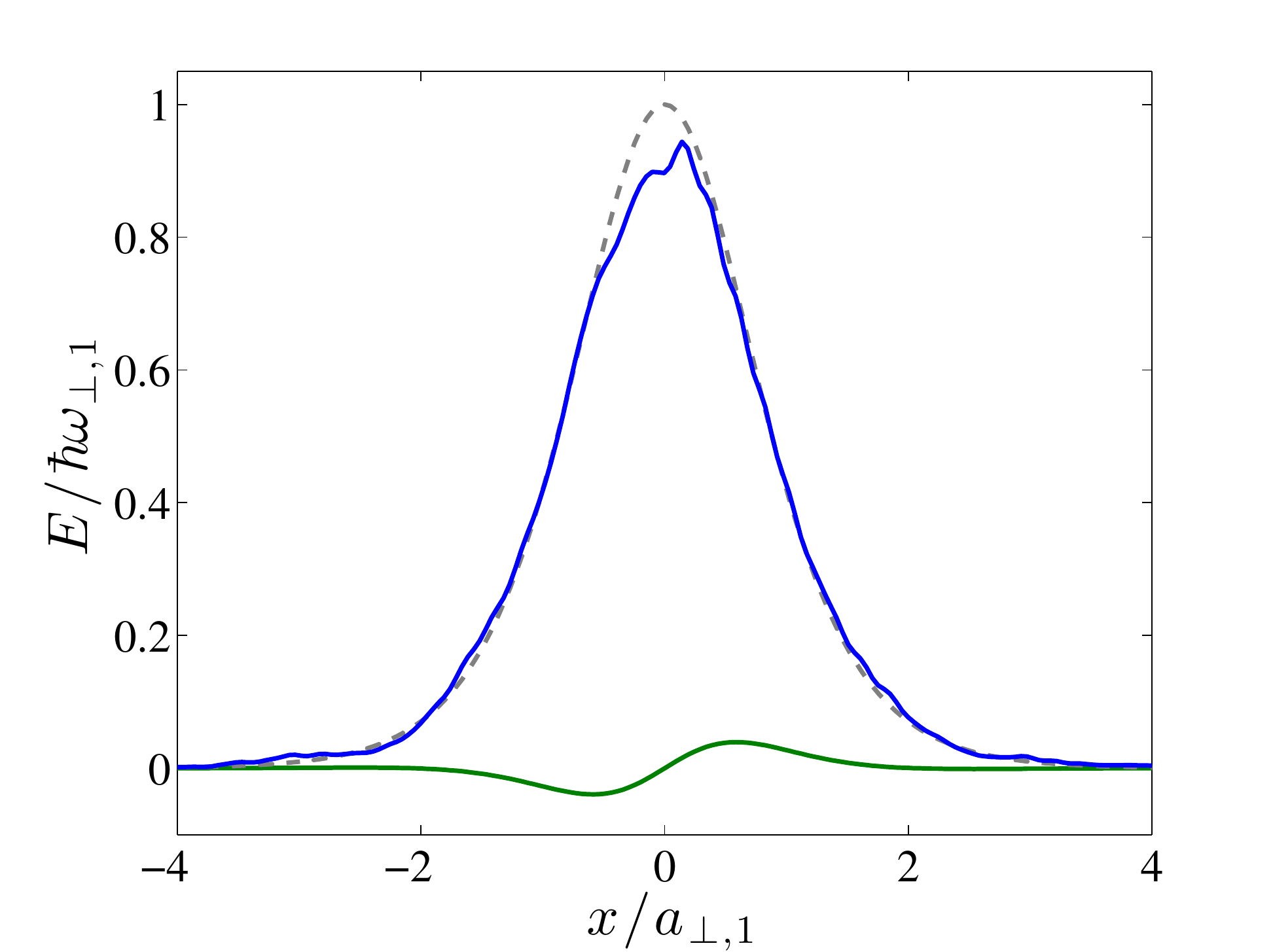}
\caption{(colour online) \rC-field particle density scaled by interaction strength $g_1n(x) = g_1 |\psi(x)|^2$ for the bright soliton analytic solution \eref{EQ:soliton}  (gray dashed) and after evolving the initial state with $\varv(0)=0.9 a_{\perp,1} \omega_{\perp,1}$ numerically for $t=300\omega_{\perp,1}^{-1}$ (blue, recentered), according to the energy-damping SPGPE \eref{1dspgpe}. The green curve shows the scattering potential $V_\ve(x,0)$ acting on the initial soliton state.
 \label{Fig:pot}}
\end{center}}
\end{figure}
\EQ{\label{1dspgpe}
(S) i\hbar d\psi(x) &={\cal P}\Big\{\Big(\left[ L +V_\ve(x,t)\right]dt-\hbar dU(x,t)\Big)\psi(x,t) \Big\}.
}
The Hamiltonian evolution is generated by $L\psi(x,t)\equiv [\mathcal{H}(x)+g_1|\psi(x,t)|^2  ]\psi(x,t)$,
where $\mathcal{H}(x)=-\hbar^2\partial_x^2/2m+V_{\rm ext}(x)$ is the single-particle Hamiltonian with external potential $V_{\rm ext}(x)$, and $g_1=2 \hbar\omega_{\perp,1} a_1$ is the 1D interaction strength of the first component with (negative) $s$-wave scattering length $a_1$ and transverse trapping frequency $\omega_{\perp,1}$. The 1D projection operator $\mathcal{P}$ implements the energy cutoff $\epsilon_{\rm cut}$ in the basis of single-particle states $\phi_n(x)$, solutions of $\mathcal{H}\phi_n(x)=\epsilon_n\phi_n(x)$. For any $F(x)$ the projector is~\cite{Davis2001b,Davis2001a,Gardiner:2003bk}
\EQ{\mathcal{P}\left\{F(x)\right\}&\equiv\int dx'\delta(x,x')F(x'),}
written in terms of the $\rC$-region delta function $\delta(x,x')=\bar\sum_n\phi_n(x)\phi^*_n(x')$, where the sum includes all modes in $C$: $\bar\sum_n\equiv\sum_{n:\epsilon_n\leq\ec}$. 
The thermal cloud $\ket{2}$, described by chemical potential $\mu$, cutoff energy $\epsilon_{\rm cut, 2}$, and temperature $T$, interacts with $\ket{1}$ via an effective potential that couples to the gradient of the matter-wave current $j(x,t)\equiv(i\hbar/2m)[(\partial_x\psi^*)\psi-\psi^*\partial_x\psi]$ with explicit form given by~\cite{Bradley:2014a,Bradley:2015cx}
\begin{subequations}
\label{allsgpe}
\EQ{
\label{ve}
V_\ve(x,t)&=-\hbar\int dx' \ve(x-x')\partial_{x'}j(x',t),\\
\label{epsdef}
\ve(x)&= \frac{ a_{12}^2}{e^{\beta|\mu|}-1}\int_{-\infty}^\infty \frac{dk\; e^{ikx}}{(2\pi (a_{\perp,1})^2)^{1/2}}{\rm erfcx}\left(\frac{|k|a_{\perp,1}}{\sqrt{2}}\right),
}
\end{subequations}
where $\beta=1/k_BT$, $a_{12}$ is the inter-component $s$-wave scattering length, $a_{\perp,1}=\sqrt{\hbar/m\omega_{\perp,1}}$ and $
{\rm erfcx}(x)\equiv e^{x^2}{\rm erfc}(x)$ 
is the scaled complementary error function~\cite{Bradley:2015cx}. The noise is a real Wiener process, with non-vanishing correlator $\langle dU(x,t)dU(x',t)\rangle=(2k_BT/\hbar)\ve(x-x')dt$. Since $\ket{2}$ is noncondensed we have taken $\epsilon_{\text{cut},2}\simeq 0$, including all of component $\ket{2}$ in the reservoir. The potential \eref{ve} damps energy from the $C$-field by opposing motion, as seen in Fig.~\ref{Fig:pot} where the right-going soliton is slowed by energy dissipation into the reservoir, causing a net drag force. The noise acts as a stochastic effective potential.
\par
The bright soliton solution \cite{Carr:2000bw} of the 1D Gross-Pitaevskii equation is
\EQ{\phi_s(x,t) =\sqrt{\frac{N_1}{2\kappa}}{\rm sech}\left(\frac{x-x(t)}{\kappa} \right) e^{i\Theta(x,t)}\label{EQ:soliton}}
where $x(t)$ is the location of the soliton centre of mass, $\varv(t)\equiv \dot{x}(t)$ is the soliton velocity, and $\Theta(x,t)=m\varv x/\hbar + \left(m\varv ^2/2 - \hbar^2/2m\kappa^2\right)t/\hbar$. The superfluid velocity $\varv_s(x)\equiv(\hbar/m)\partial_x\Theta(x)$ is spatially invariant and equal to the soliton velocity $\varv$; however the matter-wave current is spatially varying, and plays a central role in the reservoir interaction. The soliton particle number $N_1$ and the soliton width $\kappa$ are related by $N_1=2\hbar^2/(m|g_1|\kappa)$,
hence increasing the particle number results in a more spatially localised soliton. The energy per particle of the bright soliton solution is $E/N_1=m\varv^2/2- \hbar^2/6m\kappa^2$, from which we see that decreasing the soliton width $\kappa$, and thus increasing the particle number $N_1$, results in a lower energy. Our aim is to derive an equation of motion for the soliton velocity $\varv(t)$.
\par
Converting the Stratonovich stochastic differential equation (\ref{1dspgpe}) to Ito form~\cite{Gardiner:2009wp} yields
\EQ{\label{1dspgpe_Ito}
i\hbar d\psi(x,t) &={\cal P} \Big\{\mathcal{L}\psi(x,t) dt -\hbar\psi(x,t) dU(x,t)\Big\}
}
where $\mathcal{L}\psi\equiv\left[L +V_\ve -ik_BT\ve(0)\right]\psi$, 
and we use the shorthand 
\EQ{\lab{e0}
\ve(0)\psi(x)&\equiv\int dx'\; \ve(x-x')\delta(x,x')\psi(x')
}
to account for the Stratonovich correction~\footnote{An advantage of working with the projected formalism is that \eref{e0} is free from ultra-violet divergence.}. We now find an equation for the centre of mass coordinate of a system of matter waves governed by \eref{1dspgpe_Ito} by first finding a Langevin equation for the field momentum 
\EQ{P\left[\psi,\psi^*\right]&=\int dx\;\psi^*(x,t)(-i\hbar\partial_x)\psi(x,t)\label{EQ:dv}
}
via a change of variables according to Ito rules
\begin{widetext}
 \EQ{dP[\psi,\psi^*] &=\int dx \left[\frac{\bar{\delta}P[\psi,\psi^*]}{\bar{\delta}\psi(x)}d\psi(x)+\frac{k_BT}{\hbar} \int dx' \left(\frac{\bar{\delta}^{(2)}P[\psi,\psi^*]}{\bar{\delta}\psi(x)\bar{\delta}\psi^*(x')} \psi^*(x)-\frac{\bar{\delta}^{(2)}P[\psi,\psi^*]}{\bar{\delta}\psi(x)\bar{\delta}\psi(x')} \psi(x)\right)\psi(x')\ve(x-x')dt\right]+\textrm{h.c.}\quad\label{Ito}}
\end{widetext}
where all terms up to order $dt$ are retained~\footnote{The projected functional derivatives
are regular functional derivatives restricted to the \rC-region by the projector~\cite{Gardiner:2003bk}.}. Calculating the functional derivatives of \eref{EQ:dv}, and taking careful account of the Stratonovich correction, we find
\begin{subequations}
\label{EQ:const}
\EQ{
dP(t)&=F(t) dt + \sqrt{G(t)}dW(t),\\ 
F(t)&=-\int dx \;n(x,t)\partial_x  V_\ve(x,t),\\
G(t)&= 2\hbar k_B T\int dx \;dx' \; \ve(x-x')\partial_x n(x,t) \partial_{x'} n(x',t),}
\end{subequations}
with friction force $F(t)$, and noise defined by the real Wiener process $dW(t)$ with $\langle dW(t)\rangle=0$ and
$\langle dW(t)dW(t) \rangle =dt$. For $N_1$ atoms of mass $m$, the centre of mass velocity is $\varv(t)=P(t)/N_1m$, and $d\varv(t)=dP(t)/N_1m$ for number-conserving dynamics. Substituting the bright soliton wave function \eref{EQ:soliton} into \eref{EQ:const} then gives the soliton Langevin equation in Ornstein-Uhlenbeck form
\begin{subequations}
\lab{allsol}
\EQ{\label{EQ:SDE}
d\varv(t)&=-\Lambda \varv(t) dt + \sqrt{2D} dW(t),\\
\lab{lam}
\Lambda&\equiv \frac{8 \hbar N_1 a_{12}^2 }{m(\pi \kappa)^3(2\pi (a_{\perp,1})^2)^{1/2}}\frac{\mathcal{I}}{e^{\beta|\mu|}-1},\\
\lab{Idef}
\mathcal{I}&\equiv\int_{-\infty}^\infty dq\; {\rm erfcx}\left(\frac{\sqrt{2}a_{\perp,1}|q|}{\pi\kappa}\right)q^4 {\rm csch}^2(q)
}
\end{subequations}
with damping rate \eref{lam} and geometric factor \eref{Idef}.
The decay rate thus depends upon the reservoir parameters, the confinement geometry, and the form of the soliton wave function. The soliton velocity diffusion constant is $D\equiv   \Lambda k_BT/N_1m$, satisfying the fluctuation dissipation theorem. Equations (\ref{allsol}) are our main result, reducing the stochastic equation of motion for the Bose field in the Wigner representation to an Ornstein-Uhlenbeck equation for the velocity of the soliton centre of mass. In general, the energy-damping reservoir interaction stems from a quantum Brownian motion master equation~\cite{QN,Gardiner:2003bk,Bradley:2014a} for the Bose field operator describing component $\ket{1}$. Remarkably, such a reservoir interaction generates formally classical Brownian motion for the soliton velocity, with analytical damping and diffusion.
\par
A bright soliton with initial velocity $\varv(0)\equiv\varv_0$ evolves according to the formal solution
\EQ{\label{solution_vel}
\varv(t)&=\varv_0 e^{-\Lambda  t}+ \sqrt{2D}\int_0^t e^{-\Lambda\left(t-t'\right)}dW(t').
}
from which all properties of the motion may be extracted. Since $\langle dW(t)\rangle=0$, the mean velocity is $\langle \varv(t)\rangle=\varv_0 e^{-\Lambda  t}$. The steady-state variance of the velocity is
$\lim_{t\rightarrow\infty}\langle \varv^2 \rangle = D/\Lambda= k_BT/N_1m$,
a statement of the equipartition of energy for a soliton with mass $N_1m$.
The two-time correlation function of the soliton velocity 
\EQ{\lab{twotime}
\langle \varv(t)\varv(t') \rangle&=\left(\varv_0^2 -\frac{k_BT}{N_1m}\right)e^{-\Lambda(t+t')}+\frac{k_BT}{N_1m}e^{-\Lambda|t-t'|},
}
approaches the stationary form, $G_s(\tau)\equiv\lim_{t\rightarrow\infty} \langle \varv(t)\varv(t+\tau) \rangle$, given by
\EQ{G_s(\tau)=\frac{k_BT}{N_1m }e^{-\Lambda |\tau|}.\label{sstt}}
The Fourier transform then gives a Lorentzian power spectrum 
\EQ{
S_s(\omega)\equiv \frac{1}{\sqrt{2\pi}}\int d\omega \;e^{i\omega \tau}G_s(\tau)=\sqrt{\frac{2}{\pi}}\frac{k_BT}{N_1m}\frac{\Lambda}{\omega^2+\Lambda^2}.\label{spec}}
The long-term variance of the soliton position is
\EQ{
\sigma(x)^2&\equiv\langle x^2 \rangle-\langle x \rangle^2= 2\mathcal{D}t,\quad t\gg \Lambda^{-1}\label{ssxvar},}
with centre of mass diffusivity $\mathcal{D} \equiv D/\Lambda^2=k_B T/N_1m\Lambda $.
\par
\begin{figure}[t!]{
\begin{center} 
\includegraphics[width=0.9\columnwidth]{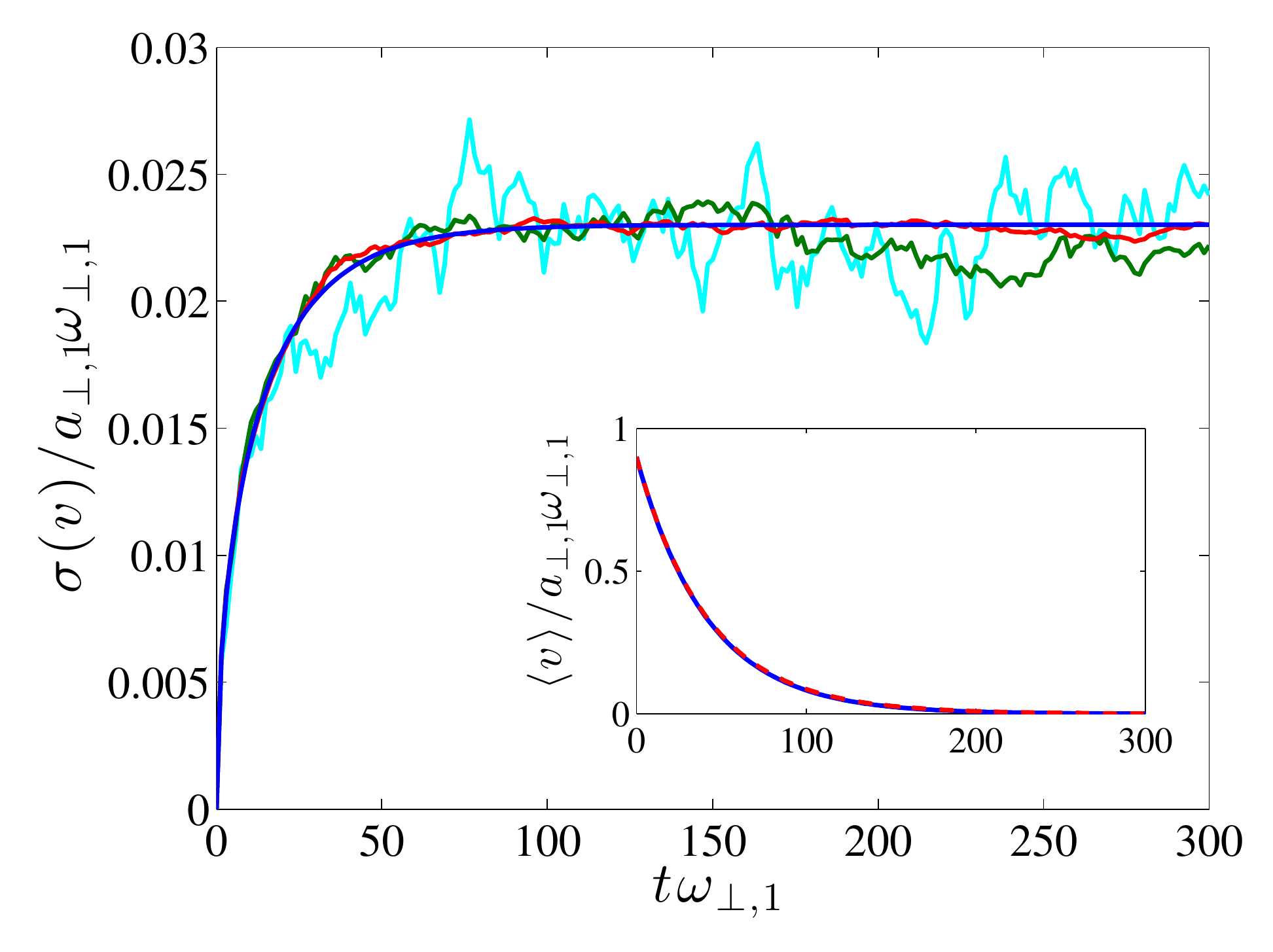}
\caption{(colour online) The variance of the soliton velocity over time for an initial position $x=0$ and initial velocity $\varv_0=0.9a_{\perp,1}\omega_{\perp,1}$. The blue line is the analytic expression derived from \eref{solution_vel} and \eref{twotime}, while the cyan, green, and red lines give the numerically obtained values from integrating \eref{1dspgpe} for 
ensembles containing 50, 500, and 5000 trajectories respectively. Inset: The average soliton velocity over time for 5000 trajectories with an initial position $x=0$ and initial velocity $\varv_0=0.9a_{\perp,1}\omega_{\perp,1}$. The solid blue line is the analytic expression derived from \eref{solution_vel}, while the dashed red line gives the values obtained from the SPGPE ensemble. 
 \label{Fig:vel}}
\end{center}}
\end{figure}
\par
We validate the analytic solution by numerically integrating the SPGPE (\ref{1dspgpe}) using the semi-implicit Euler method \cite{Rooney:2012gb,Rooney:2014kc}, on a grid consisting of $M=1024$ points with periodic boundary conditions. The transverse trapping is taken to be harmonic with frequency $\omega_{\perp,1}=2\pi\times 200$Hz, giving a quasi-1D degenerate component $\ket{1}$~\cite{Bradley:2015cx}. The length of the toroid is set to $L=50a_\perp=  38.6\mu {\rm m}$, and the $s$-wave scattering lengths are $a_1=-1.5a_0$, $a_{12}=88a_0$ with $a_0$ the Bohr radius. The initial condition is given by the wave function (\ref{EQ:soliton}), with soliton width $\kappa=a_{\perp,1}=0.77\mu{\rm m}$. This gives a soliton containing $N_1= 10151$ $^{85}$Rb atoms, well within the experimentally accessible range \cite{Khaykovich:2002dk,Nguyen:2014it,McDonald:2014fg}. The temperature is held constant at $T=51.6 {\rm nK}\simeq 2T_{c,2}$ where $T_{c,2}=26.3 {\rm nK}$ is the transition temperature for a 3D gas of $N_2=1.5\times 10^4$ $^{87}$Rb atoms in a toroid of length $L$, with transverse harmonic trapping scale $a_{\perp,2}=1.97\mu{\rm m}$~\cite{Bradley:2009bd}. The transverse thermal scale is $a_{T,2}=(k_B T/m_2\omega_{\perp,2}^2)^{1/2}=11.7\mu{\rm m}$, and so $a_{\perp,2}\ll a_{T,2}$ ensuring a 3D thermal cloud.
Our parameters give a centre of mass diffusivity of $\mathcal{D}=17.1 \mu{\rm m}^2{\rm s}^{-1}$, and a characteristic decay time of $\Lambda^{-1}=0.0338$s. After $t\sim 1{\rm s}$ the soliton travels an rms distance $\sigma(x)\simeq 5.8\mu {\rm m}\simeq 7.6\kappa$, providing a measurable  signature of Brownian motion, accessible within typical condensate lifetimes~\cite{Altin:2010kb}. 
\par
\begin{figure}[t!]{
\begin{center} 
\includegraphics[width=\columnwidth]{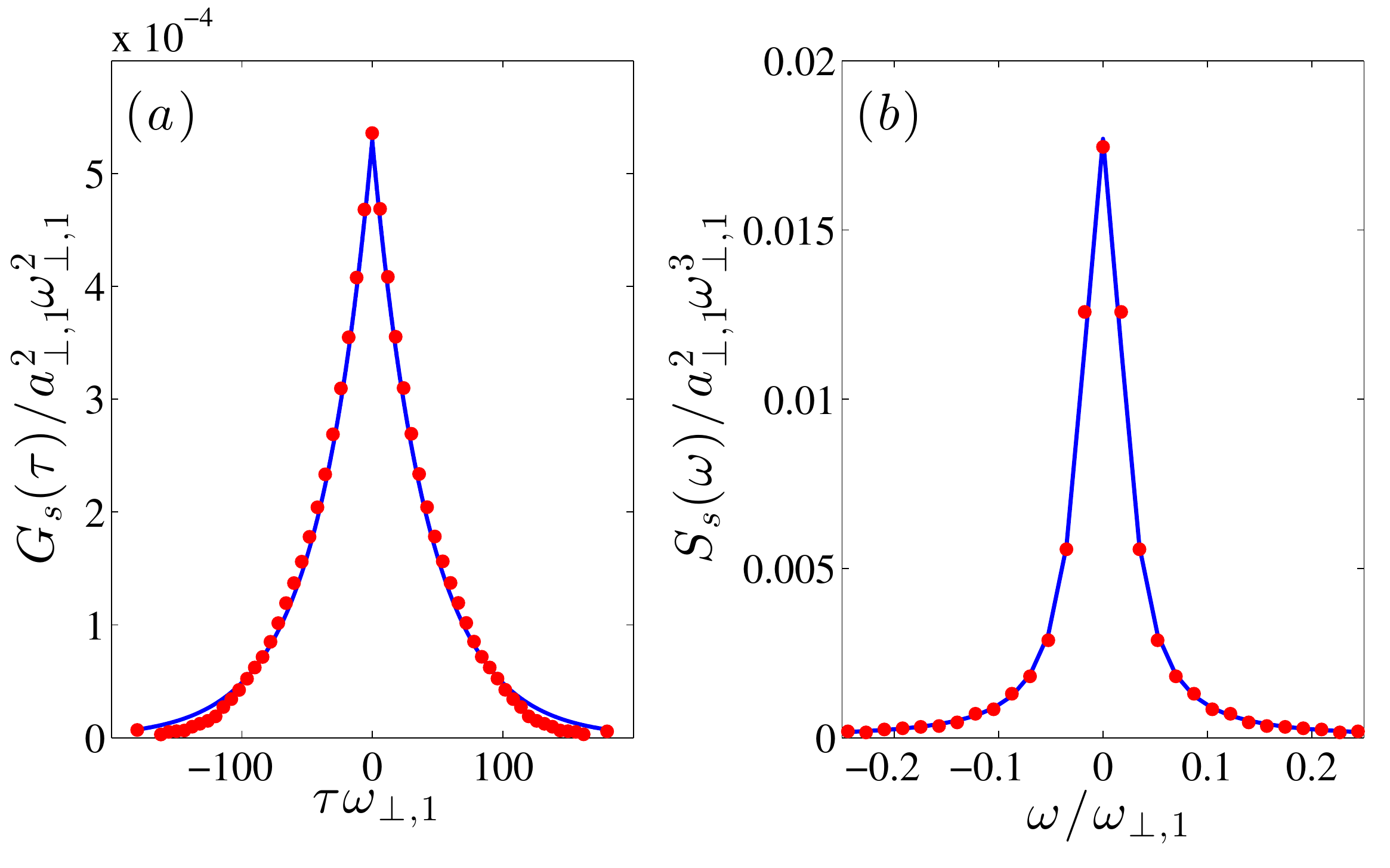}
\caption{(colour online) The steady-state (a) two-time correlation function and (b) spectrum of the soliton velocity constructed using 500 trajectories of an initially stationary soliton at $x=0$. The solid blue lines are the analytic expressions \eref{sstt} and \eref{spec} respectively, while the red points show the values obtained numerically from the SPGPE.
\label{Fig:twotime}}
\end{center}}
\end{figure}

\fref{Fig:vel} shows the rms velocity $\sigma(\varv)$ over time for several ensembles of bright soliton evolution from initial velocity  $\varv_0=0.9a_{\perp,1}\omega_{\perp,1}=0.97{\rm mms}^{-1}$, compared with the prediction from \eref{twotime}. For all ensembles, convergence to the analytical curve is apparent with increasing ensemble size. The inset of \fref{Fig:vel} shows the ensemble-average velocity $\langle \varv \rangle$ of the bright soliton dynamics predicted by the SPGPE, compared with the analytical result from \eref{solution_vel}. \fref{Fig:twotime} shows the steady-state two-time correlation function and the power spectrum of the soliton velocity, both analytically and numerically. For all the quantities we measure the numerical data shows excellent quantitative agreement with the predictions of Eq.~\eref{EQ:SDE}. Experiments have also been performed for a $^{85}$Rb-$^{87}$Rb mixture exhibiting soliton-like evolution for a system that is outside the 1D regime~\cite{McDonald:2014fg}. A variational ansatz could be used to extend the present approach to that regime, in the presence of a distinct 3D thermal cloud.
\par
We have studied the dissipative evolution of a matter-wave bright soliton immersed in a thermal cloud of a distinct atomic species using the stochastic projected Gross-Pitaevskii equation~\cite{Gardiner:2003bk,Bradley:2014a,Bradley:2015cx}. Number-conserving dissipation of kinetic energy to the thermal cloud induces Brownian motion of the soliton centre of mass, a clear signature of the system-reservoir interaction for a Bose-Einstein condensate embedded in a distinguishable thermal cloud. Analytical expressions for the drift and diffusion constants characterising the bright soliton motion reveal its dependence upon the reservoir parameters, confining geometry, and upon the shape of the bright soliton. Predictions of the soliton Langevin equation are in close agreement with the evolution of the matter-wave field according to the full stochastic projected Gross-Pitaevskii equation, demonstrating soliton diffusion that should be accessible with current experimental techniques, and suggesting a route for direct experimental measurements of the energy-damping reservoir interaction. Clear observation of soliton Brownian motion would have fundamental implications for the theory of open quantum systems~\cite{Gardiner:2003bk,Rooney:2012gb,Bradley:2014a,Bradley:2015cx,Daley:2014hc}, vortex dynamics~\cite{Fedichev1999,Schmidt2003,Duine2004,Bradley:2008gq,Rooney:2010dp,Rooney:2011fm} and quantum turbulence~\cite{Henn09a,Neely:2013ef}, and the dynamics of the BEC phase transition~\cite{Weiler:2008eu,Damski10a,Su:2013dh,Lamporesi:2013bi,Navon:2015jd,McDonald:2015ju}. A deeper understanding of energy damping may also reveal new routes to quantum degenerate matter ~\cite{Myatt:1997ct,Zwierlein:2005kl}.
\acknowledgements
We thank Sam Rooney, Blair Blakie, and Niels Kj\ae rgaard for stimulating discussions. ASB is supported by a Rutherford Discovery Fellowship administered by the Royal Society of New Zealand, and by the Dodd-Walls Centre for Photonic and Quantum Technologies.
%

\end{document}